\documentstyle[tighten,preprint,eqsecnum,aps,prd]{revtex}
\begin{document}
\draft

\hyphenation{
mani-fold
Schwarz-schild}


\preprint{\vbox{\baselineskip=12pt
\rightline{PP98--70}
\rightline{gr-qc/9712082}}}
\title{Light bending and perihelion precession: A unified approach}
\author{Dieter R. Brill\footnote{Electronic address:
brill@physics.umd.edu}
and Deepak Goel\footnote{Electronic address: dgoel@wam.umd.edu}}
\address{Department of Physics,
University of Maryland,
College Park, MD 20742-4112}
\date{\today}
\maketitle
\begin{abstract}
The standard General Relativity results for precession of particle orbits and 
for bending of
null rays are derived as special cases of perturbation of a quantity that is
conserved in Newtonian physics, the Runge-Lenz vector. First this method is
applied to give a derivation of these General Relativity effects
for the case of the spherically symmetric Schwarzschild geometry. Then the
lowest order correction due to an angular momentum of the central body is
considered. The results obtained are well known, but the method used is rather
more efficient than that found in the standard texts, and it provides a good
occasion to use the Runge-Lenz vector beyond its standard applications in
Newtonian physics.
\end{abstract}
\narrowtext

\section{Introduction}
\label{sec:intro}

Light bending and perihelion precession are the two most important effects 
on orbits caused by the General Relativity corrections to the 
Newtonian gravitational field of the sun. The standard
derivation treats these two effects in different ways, without any apparent
connection between them. Yet, in the usual Schwarzschild coordinates they
are both due to the same, single relativistic correction to the Newtonian
potential, so it is of some interest to use the same method to derive either
effect.

The key to the present unified treatment is the Runge-Lenz vector.  In
Newtonian physics, where the two effects are absent, this vector is
constant and points from the center of attraction to the orbit's
perihelion.\footnote{Although our treatment is not confined to the solar
system, we use this term to denote the point of closest approach to the
center because it seems more familiar (and more etymologically consistent)
than the more correct term, pericenter.} Its non-constancy in General
Relativity therefore is a measure of either effect. The Runge-Lenz vector
was established as a useful tool by 1924 at the latest, but it did
not become popular until the 1960's.\footnote{For a history of
the Laplace-Runge-Lenz vector, see ref.~\cite{GS}.}

Since then a number of papers that exploit its advantages have graced the
pages of this Journal (see ref.~\cite{AB} and the references cited
therein), and the results to be reported here can in essence be found in
earlier papers, but the unified viewpoint vis a vis General Relativity is
perhaps new. In addition the ``magnetic" gravitational effects due to a
rotating central body are treated here with this method. 

\section{General Relativistic Equations of Motion (no rotation)}
\label{sec:EOM}
The motion to be considered is that of a ``test particle" that moves
along a geodesic in the spacetime exterior to the central body. If this body is 
non-rotating, spherically symmetric, and has total mass $M$,
the exterior spacetime geometry is described by the Schwarzschild line element 
\begin{equation}
ds^2 = -\left(1 - {2M\over r}\right) dt^2 + {dr^2\over 1 - {2M\over r}} + 
r^2 d\Omega^2.
\label{SS}
\end{equation}
Here the coordinates $t, r, \theta$ and $\phi$ are one of many equally valid
choices for labeling spacetime points, but they can nevertheless be invariantly
characterized \cite{MTW}. (For example, $4\pi r^2$ is the area of the sphere
$r =$ const, $t =$ const, and $\partial/\partial t$ is a timelike Killing
vector.) The geodesic law of motion is essentially equivalent to the
conservation laws that follow from the symmetries of the geometry and the
conservation of rest mass. Because of the principle of equivalence we may
assume without loss of generality that the rest mass is $\epsilon$, where
$\epsilon = 1$ for particles of finite rest
mass, and $\epsilon = 0$ for light (photons).
The other conserved quantities correspond to angular momentum ${\bf L}$, and
energy $E$. We use $\tau$ to denote the proper time along the 
geodesic.\footnote{For the case of light, $\tau$ is an affine parameter that 
is defined only up to scale transformations. The quantities $E$ and $L$
are therefore similarly defined only up to such re-scaling. The final,
physical results will contain only ratios of such quantities.} As usual
we can choose the vector $\bf L$ to be normal to the plane $\theta = \pi/2$
and we then have\footnote{For a derivation, see references \cite{MTW,SCH}
or section \ref{sec:ROT} below.}
\begin{eqnarray}
E & = & \left(1 - {2M\over r}\right) {dt\over d\tau} \\
L & = & r^2 {d\phi\over d\tau} \\
{\cal E} \equiv {1\over 2}(E^2 - \epsilon) & = & {1\over 2}\left(dr\over d\tau \right)^2 - 
{\epsilon M\over r} + 
{L^2\over 2r^2} - {ML^2\over r^3}\, . 
\label{non-Newt}
\end{eqnarray}
Equation (2.4) has the form of conservation of energy in an effective potential. 
Except for the presence of $\tau$ instead of
$t$, Eqs (2.2 - 2.4) are the same as the Newtonian equations of 
motion of a particle of unit mass and total energy $\cal E$ in 
a potential $V = -\epsilon M/r - ML^2/r^3$.
Thus for particles as well as for light 
the relativistic motion in proper time $\tau$ is the same as the Newtonian motion
in Newtonian time $t$ if the potential is modified by the single term 
$-ML^2/r^3$.
We note that no slow motion assumption or other approximation is involved in
this correspondence. If we are only interested in the orbit equation, then the
difference between $t$ and $\tau$ does not matter, because either one will be
eliminated in the same way in favor of $\phi$ via Eq (2.3). 

\section{Secular Change of Orbits}
\label{Orbits}
We treat the modification $ML^2/r^3$ of the Newtonian potential as a 
perturbation\footnote{This means that $M/r \ll 1$, where $r$ is a typical
orbit radius, which follows if we assume $M/L \ll 1$}
and compute the consequent changes in direction of the Runge-Lenz vector,
defined by
\begin{equation}
{\bf A} = {\bf v} \times {\bf L} - \epsilon M{\bf e_r}
\label{RL}
\end{equation}
where all bold-face quantities are 3-vectors, and $\bf e_r$ denotes a unit 
vector in the {\bf r}-direction. The time parameter is $\tau$ as above, so 
that, for example, ${\bf v} = d{\bf r}/d\tau$. The rate of change of {\bf A}
is (as, for example, in ref.~\cite{GST})
\begin{equation}
{d{\bf A}\over d\tau}  = \left(r^2{\partial V\over\partial r} - 
{\epsilon M} \right) {d{\bf e_r}\over d\tau}= 
\left({3ML^2\over r^2}\right){d\phi\over d\tau}\,{\bf e_\phi}\,.
\end{equation}
The direction of {\bf A} therefore changes with angular velocity
\begin{equation}
\omega\hskip-7pt\omega = {\bf A \times \dot A\over A^2} =
\left({3ML^2\over A^2r^2}\right){d\phi\over d\tau}\,
{\bf A}\times {\bf e_\phi} 
\label{omega}
\end{equation}
and its total change when the particle moves from $\phi_1$ to $\phi_2$ is
(assuming this change is small, and that {\bf A} is originally in the 
$\phi = 0$ direction)
\begin{equation}
\Delta\alpha = \int_{\phi_1}^{\phi_2} \omega\, d\tau = 
3ML^2\int_{\phi_1}^{\phi_2}{\cos\phi\,d\phi\over Ar^2}\,.
\end{equation}

When {\bf A} is constant, and in the direction $\phi = 0$, we have
\begin{equation}
{\bf A}\cdot {\bf r} = Ar\cos\phi = L^2 - \epsilon M r .
\label{ellipse}
\end{equation}
The bound orbits ($\epsilon = 1$) are therefore ellipses with eccentricity 
$e = A/M$ and semi-major axis $a = {L^2\over M(1-e^2)}$. For the unbound, 
straight orbits of light rays ($\epsilon = 0$), $L^2/A \equiv b$ 
is the impact parameter, and because these orbits are traversed
at the speed of light we have $L/E = b$. In either case {\bf A} points from 
the center to the perihelion.
When {\bf A} changes slowly the orbits still have this approximate
shape, but their orientation and shape will change slowly. We calculate
the lowest order changes due to the General Relativistic correction in 
$V_{\rm eff}$ by substituting the unperturbed orbit (3.5) into Eq (3.4):
\begin{equation}
\Delta\alpha = {3M \over AL^2} \int_{\phi_1}^{\phi_2} 
(A\cos\phi + \epsilon M)^2\cos\phi\, d\phi.
\label{delal}
\end{equation}

\subsection{Perihelion precession}
For a particle in a bound orbit it is customary to find the angular change
of the perihelion during one revolution (in $\phi$) of the particle. Because
{\bf A} points to the perihelion, this angle is given by Eq (3.6), when
$\phi_2 - \phi_1 = 2\pi$. 
\begin{equation}
\Delta\alpha = {3M \over L^2} \int_0^{2\pi} {(A\cos\phi + M)^2\over A}
\cos\phi\, d\phi = {6\pi M^2\over L^2} = {6\pi M\over a(1-e^2)}\,.
\end{equation}
This is the usual perihelion formula.

\subsection{Light bending}
Here also the deflection is given by $\Delta\alpha$ of Eq (3.6), but $\phi$
changes from $-\pi/2$ to $\pi/2$ with respect to the perihelion (and of course
$\epsilon = 0$):
\begin{equation}
\Delta\alpha = {3M \over L^2}\int_{-\pi/2}^{\pi/2}  
A\cos^3\phi\, d\phi = {4MA\over L^2} = {4M\over b}\,.
\end{equation}
This is the usual light deflection formula.

That the light deflection should follow from the same, $O(1/r^3)$
correction to the Newtonian effective potential as the perihelion rotation
may be somewhat surprising, because the deflection is frequently
heuristically explained as an action of the Newtonian $O(1/r)$ potential
on light. Indeed, an effective potential for $dr/dt$ would contain
$O(1/r)$ terms, but in the present choice of variables these are absent
--- illustrating once again the arbitrary nature of coordinates in a
generally covariant theory. 

\section{Slowly rotating central bodies} 
\label{sec:ROT} 
If the body is
slowly rotating\footnote{We work only to first order in $J$;  more
precisely, we assume $J$ \tiny $\stackrel{\textstyle <}{\sim}$ 
\footnotesize $ ML$ and, as before, $M/r\sim M^2/L^2 \sim \varepsilon \ll 1$,
so that $J/r^2 \sim \varepsilon^{3/2}$.}
in the $\phi$-direction
with angular momentum $J$, the metric (2.1) is modified by the
Lense-Thirring term \cite{LT}, $- {4J\over r}\sin^2\theta d\phi dt$ and by
a quadrupole term that describes the distortion of the body.  The
quadrupole term can be treated in the same way as the relativistic
correction, and will therefore not be further considered. The
Lense-Thirring term breaks the spherical symmetry, so on symmetry grounds
only $p_\phi$, the angular momentum conjugate to $\phi$ (called $L$
below), is conserved. Nonetheless the ``total angular momentum" $Q^2 =
p_\theta^2 + \cot^2\theta\,p_\phi^2$, where $p_\theta$ is the momentum
conjugate to $\theta$, is also conserved to first order in $J$. Thus the
entire motion can be formulated in terms of conserved quantities, and one
finds that the Newtonian angular momentum {\bf L} precesses around the
$z$-direction.\footnote{The integration of the equations of motion for
orbits of general orientation was one of the aims of the Lense and
Thirring paper \cite{LT}.  For a treatment using the Runge-Lenz vector,
see references \cite{LL}.  For a summary of all the relativistic effects
on orbits see reference \cite{CW}.  For the geometrical reason for the
conservation of $Q^2$ see reference \cite{BP}.} However, we will confine
attention to the case when $\bf J$ and $\bf L$ are parallel, the motion is
confined to the equatorial plane, and the general relativistic correction
is completely described by the behavior of {\bf A}. 

\subsection{Equations of Motion and Runge-Lenz vector} 
A Lagrangian $\cal L$ for a particle or a light beam moving the in equatorial 
plane of metric
\begin{equation}
ds^2 = -\left(1 - {2M\over r}\right) dt^2 + {dr^2\over 1 - {2M\over r}} + 
r^2 d\Omega^2- {4J\over r}\sin^2\theta d\phi dt
\label{LT}
\end{equation}
is given by
\begin{equation} 2{\cal L} = -\left(1-{2M\over r}\right)\dot t^2 + {\dot
r^2\over 1-{2M\over r}} + r^2 \dot\phi^2 - {4J\over r}\dot\phi \dot t\,. 
\label{ell} 
\end{equation} 
Here the dot indicates differentiation with
respect to proper time; this implies that $\cal L$ is constant,
\begin{equation} 2{\cal L} = -\epsilon\,.  
\label{norm} 
\end{equation}
Since $\cal L$ is independent of $t$ and $\phi$ we have the conserved
quantities, 
\begin{equation} 
-E = {\partial{\cal L}\over\partial\dot t} =
-\left(1-{2M\over r}\right)\dot t - {2J\over r}\dot\phi \qquad \qquad L =
{\partial{\cal L}\over\partial\dot\phi} = r^2\dot\phi - {2J\over r}\dot t\,. 
\end{equation} 
We solve for $\dot t$ and $\dot \phi$ to first order in $J$, 
\begin{equation} 
\dot\phi = {L\over r^2} + {2JE\over r^3} \qquad
\dot t = {E\over 1-{2M\over r}} - {2JL\over r^3} 
\label{t} 
\end{equation}
and substitute into Eq (\ref{norm}) to obtain a ``conservation of energy"
in an effective potential, 
\begin{equation} 
E^2-\epsilon = \dot r^2 -
{2\epsilon M\over r} + {L^2\over r^2} -{2ML^2\over r^3} + {4JLE\over
r^3}\,.  
\label{r} 
\end{equation} 
The effective potential in this equation
contains two non-Newtonian terms.  The first was already encountered in Eq
(\ref{non-Newt}) and causes the ``standard" relativistic correction; the
second is due to the Lense-Thirring addition to the metric.\footnote{With
our assumptions as spelled out in footnote 6 the
Newtonian terms of the effective potential are of order $\varepsilon$,
both non-Newtonian terms are of order $\varepsilon^2$, and typical terms that
are neglected are $J^2E^2/r^4 \sim JEML/r^4 \sim \varepsilon^3, \quad
J^2L^2/r^6 \sim \varepsilon^4$ etc.}

Because in the rotating case there is a difference between kinematic
angular momentum ($r^2\dot\phi$) and canonical angular momentum $L$, the
Runge-Lenz vector {\bf A} can be defined in various ways, but there is
no difference in the precession rate one calculates from them. A convenient
choice is 
\begin{equation}
{\bf A} = {\bf v}\times\left({\bf L} - {2{\bf J}E\over r}\right)  -\epsilon M{\bf e_r}
=\left({L^2\over r} - \epsilon M\right){\bf e_r}
- \dot r \left(L - {2JE\over r}\right){\bf e_\phi}
\end{equation}
because it simplifies the equation of motion for {\bf A}, and still gives 
elliptical orbits for any J when {\bf A} is constant,
\begin{equation}
{\bf A \cdot e_r} = A\cos{\phi} ={L^2\over r} -\epsilon M\,. 
\label{orb}
\end{equation}
The equation of motion for {\bf A} can be derived from Eqs (\ref{t}, \ref{r}):
\begin{equation}
{\bf \dot A}  = \left({3ML^2\over r^2} - {8\epsilon MJE\over Lr} +
{2JE(\epsilon -E^2)\over L}\right)\dot\phi\, {\bf e_\phi}\,.
\label{Adot}
\end{equation}
By substituting Eq (\ref{Adot}) into Eq (\ref{omega}) and integrating, 
using (\ref{orb}) for $1/r$, we now find that the total change in {\bf A} 
when the particle moves from $\phi_1$ to $\phi_2$ is
\begin{equation}
\Delta\alpha =  \int_{\phi_1}^{\phi_2}\left({3M\over AL^2}(A\cos\phi + \epsilon M)^2
-{8\epsilon MJE\over AL^3}(A\cos\phi+\epsilon M) + {2JE(\epsilon -E^2)\over AL}
\right)\cos\phi\, d\phi\,.
\label{gen}
\end{equation}

\subsection{Perihelion motion}

To obtain the perihelion motion we evaluate Equation (\ref{gen}) over one
revolution ($\phi_1=0,\, \phi_2=2\pi$) with $\epsilon = 1$. Since the
particle-velocity is nonrelativistic, we may set $E=1$ to the lowest 
order:\footnote{Formally, this follows from the $M^2 \ll L^2$ assumption and requiring 
bound orbits.}

\begin{equation}
\Delta\alpha = {6\pi M^2\over L^2} - {8\pi J M E\over L^3} = 
{6\pi M\over a(1-e^2)} - {8\pi J\over 
M^{1/2}\left(a(1-e^2)\right)^{3/2}}\,.
\label{Jprecess}
\end{equation}
The first term is the ``standard" general relativistic precession already found in
Section III, and the second term is due to the rotation of the central 
body.\footnote{The last term in Eq (\ref{Jprecess}) changes sign if $L$ is
antiparallel to $J$. Both terms are of order $\varepsilon$.}
For nearly circular orbits we can interpret this second term as due to two 
causes: one is the rotation in $\phi$ of the ``locally non-rotating observer"
that makes the
Lense-Thirring term of Eq (\ref{ell}) disappear at the radius of the particle,
an amount $4\pi J/aL$; the other is the ``differential rotation" due to the
$1/r^3$ fall-off of the second non-Newtonian term
in the effective potential of Eq (\ref{r}). This 
contribution causes precession by an amount $-12\pi J/aL$, in the same way as 
the first non-Newtonian term
causes the ``standard" precession.\footnote{For non-equatorial orbits the first
contribution is a precession about {\bf J}, whereas the second contribution is
a precession about {\bf L}, proportional to $\bf J \cdot L$.}

\subsection{Light bending}
For the effect on light we put $\epsilon = 0$ in Eq (\ref{gen}) and 
integrate from $\phi = -\pi/2$ to $\pi/2$ as in section \ref{Orbits} B,
\begin{equation}
\Delta\alpha = {3MA\over L^2}\cdot{4\over 3} - 
{2JE^3\over AL}\cdot 2 = {4M\over b}- {4J\over b^2}\,.
\end{equation}
The effect of the angular momentum $J$ on both precession and bending is
negative.\footnote{However, the relative the contribution of $J$ to
light bending is less: we have $M/b \sim \varepsilon$, but $J/b^2 \sim
\varepsilon^{3/2}$.}  This is the same ``differential" dragging effect that makes
a gyroscope in the equatorial plane precess in the opposite direction to
the central body's rotation \cite{CW}.

\section{Conclusions}
For non-rotating spherically symmetric central masses
we have seen that the two important general relativistic corrections to the
Newtonian gravitational motion, namely perihelion precession and light bending, 
follow from the same correction term in the effective radial potential; and
that either effect can be viewed as change in the Runge-Lenz vector associated
with the orbit. Because both effects follow by simple evaluation from one
formula (Eqs \ref{delal}, \ref{gen}), the effort is only about half of the
usual procedure; moreover it gives occasion to review and apply the 
Runge-Lenz vector. We have shown the extension of this calculation to
equatorial orbits of a rotating body; 
relativistic corrections to parabolic and hyperbolic orbits can similarly 
be evaluated by this method.

\end{document}